\documentclass{IEEEtran}
\usepackage{ amsmath, amssymb, multicol, graphicx, gensymb}
\usepackage{flushend}

\title{Sterilization of Bacteria with Low-Pressure Low-Temperature Plasma Discharge}

\author{\IEEEauthorblockN{Angel Gonz\'alez-Lizardo\IEEEauthorrefmark{1} , Jairo Rond\'on\IEEEauthorrefmark{2}}%
\bigskip
\IEEEauthorblockA{\IEEEauthorrefmark{1}Polytechnic University of Puerto Rico,\\
agonzalez@pupr.edu}
\IEEEauthorblockA{\IEEEauthorrefmark{2}Polytechnic University of Puerto Rico,\\
jrondon@pupr.edu}
}

\begin{document}
\maketitle

\begin{abstract}
The study investigates the effectiveness of using low-pressure oxygen plasma, generated by microwave discharge, to sterilize bacterial spores of \emph{Bacillus Stearothermophilus} and \emph{Bacillus Subtilis}. Plasma, a highly ionized state of matter consisting of positive ions and free electrons, is created by applying high temperatures or accelerating electrons through an electric field. This method provides an innovative alternative to traditional sterilization techniques, particularly beneficial for heatsensitive materials. In the experiments, microwave at 2.45 GHz with 1000 W and 500 W  intensities (990 W and 495 W) were applied in a vacuum chamber at a pressure of $10^{-3}$ Torr. Spores were exposed to the plasma using two types of holders: Petri dishes and centrifuge tubes, for 1, 5, and 10 minutes. The findings revealed that direct exposure to oxygen plasma at a 1000 W microwave power effectively inactivated spores in Petri dishes, especially for \emph{B. Subtilis}. Conversely, centrifuge tubes did not allow sufficient plasma exposure, leading to the survival of bacteria posttreatment. The inactivation process involves UV induced DNA damage and the erosion of spores' protective layers by oxygen radicals. The study concludes that low-pressure oxygen plasma is a promising technology for sterilizing heat-sensitive materials. Optimal conditions identified include a 1000 W microwave power and exposure times of at least 5 minutes for \emph{B. Stearothermophilus} and 1 minute for \emph{B. Subtilis}. Petri dishes proved more effective than centrifuge tubes due to the closer and more direct plasma contact. This method offers a safe and efficient alternative to traditional sterilization techniques, with significant potential for applications in the sterilization of medical devices and other delicate materials.
\end{abstract}

\section{Introduction}

Plasma is a state of matter that is heated beyond its gaseous state to such a high temperature that atoms lose at least one electron from their outer shells, leaving behind positive ions in a sea of free electrons. Plasma can also be created by heating a neutral gas to very high temperatures. This usually requires the application of high temperature externally by injecting highvelocity ions or electrons that collide with the gas particles, increasing their thermal energy. The gas electrons are also accelerated by an external electric field \cite{goldston2020introduction}. Sterilization is a fundamental process in various industries, especially in biomedical engineering, where the complete elimination of microorganisms is crucial to ensure the safety and effectiveness of equipment and procedures \cite{kumar2021basic}. Although several traditional sterilization methods exist, such as autoclaving and ultraviolet radiation, these have limitations, particularly in the sterilization of heatsensitive materials and penetration into hardtoreach areas. Plasma, the fourth state of matter, offers an innovative and effective solution to these challenges. Plasma forms when a gas is heated to the point where atoms lose at least one electron, creating a medium of positive ions and free electrons \cite{nishikawa2000plasma}, \cite{gonzalez2024water}. This ionized state can be generated through the injection of highvelocity ions or electrons that collide with gas particles, increasing their thermal energy \cite{chen2012introduction}. Additionally, electron acceleration in the gas can be induced by an external electric field. Plasma sterilization uses this highly reactive medium to deactivate and eliminate microorganisms, leveraging various mechanisms, including free radicals and ultraviolet radiation, which damage essential cellular components like DNA and proteins \cite{rutala2019guideline}. Unlike traditional methods, plasma can penetrate complex surfaces and protected areas, making it particularly effective in disinfecting medical and electronic devices with intricate geometries.

\section{Experimental Procedure}
\begin{figure}[h]
  \centering
  \includegraphics[width=.9\columnwidth]{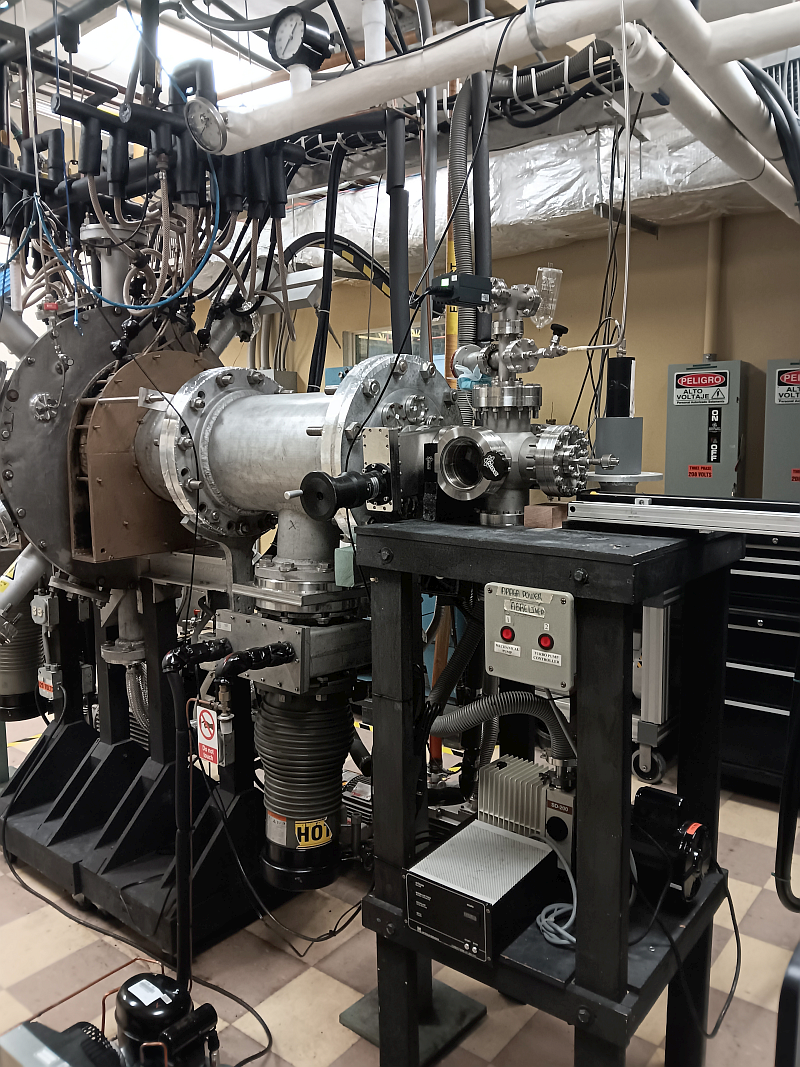}
  \caption{PUPR Plasma Laboratory}\label{F1}
\end{figure}

\begin{figure}
  \centering
  \includegraphics[width=.8\columnwidth]{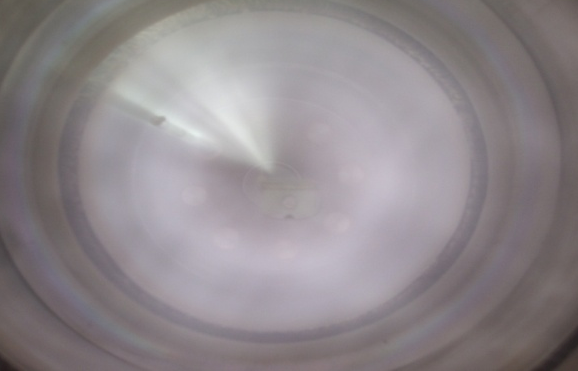}
  \caption{Oxygen plasma}\label{F2}
\end{figure}

\begin{figure}
  \centering
  \includegraphics[width=\columnwidth]{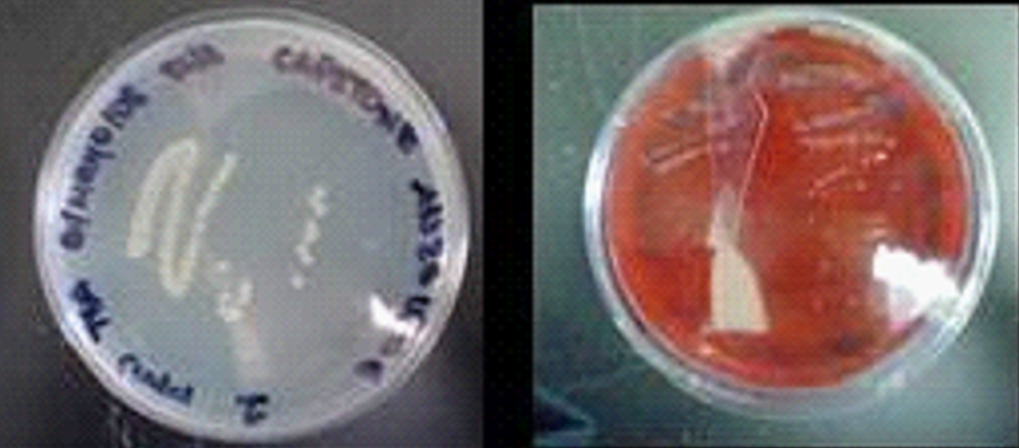}
  \caption{Bacteria culture a) \emph{B. Stearothermophilus} b) \emph{B. Subtilis}}\label{F3}
\end{figure}

The microwave power of 2.45 GHz was changed to 1000 W microwave and 500 W  microwave to 990 W and 495 W respectively. The vacuum chamber (Figure \ref{F1}) was pumped down to $10^{-3}$ Torr with a turbo-molecular pump. For the experiments of plasma sterilization, we used oxygen discharges at the pressure of $10^{-3}$ Torr (Figure \ref{F2}). As the biological indicators, we used \emph{Bacillus Stearothermophilus} and \emph{Bacillus Subtilis}. After plasma irradiation, the spore samples are incubated for 24 hours at 55°C to 65°C for \emph{B. Stearothermophilus} and \emph{B. Subtilis} at 24°C at the same 24hours (Figure \ref{F3}). 

We used 2 types of holders to expose bacteria in the plasma: the Petri dish and the centrifuge tube. These 2 were exposed in the plasma at 1000 W microwave and the petri dish at 10 for different time of 10 minutes, 5 minutes, and 1 minute as shown in Table \ref{T1} and Table \ref{T2}. The reason for this was to find a better exposition of plasma in the plasma machine. Given the plasma machine centrifuge tube did not get the exposition we needed.

\begin{table*}[!ht]
  \centering
  \caption{Tube samples results}\label{T1}
  \renewcommand{\arraystretch}{1.2}
  \begin{tabular}{|c|c|c|c|c|c|c|}
\hline
\multicolumn{2}{|c|}{Exposure Time 10 minutes} & Pressure & Temperature (K) & Density (cm$^{-3}$) & Microwave power & Survivability of spores\\
\hline
Sample & Bacteria & \multicolumn{5}{|c|}{}\\
\hline
1 & \emph{B. Subtilis} & $2.20 \times 10^{-3}$ & 19871.75 & $4.32\times 10^{9}$ & 1000 W & Bacterial Presence\\
\hline
2 & \emph{B. Subtilis} & $2.20 \times 10^{-3}$ & 19871.75 & $4.32\times 10^{9}$ & 1000 W & Bacterial Presence\\
\hline
3 & \emph{B. Subtilis} & $2.20 \times 10^{-3}$ & 19871.75 & $4.32\times 10^{9}$ & 1000 W & Bacterial Presence\\
\hline
4 & \emph{B. Subtilis} & $2.20 \times 10^{-3}$ & 19871.75 & $4.32\times 10^{9}$ & 1000 W & Bacterial Presence\\
\hline
5 & S. Thermo & $1.20 \times 10^{-3}$ & 23288.43 & $4.63\times 10^{9}$ & 1000 W & Bacterial Presence\\
\hline
6 & S. Thermo & $1.20 \times 10^{-3}$ & 23288.43 & $4.63\times 10^{9}$ & 1000 W & Bacterial Presence\\
\hline
7 & S. Thermo & $1.20 \times 10^{-3}$ & 23288.43 & $4.63\times 10^{9}$ & 1000 W & Bacterial Presence\\
\hline
8 & S. Thermo & $1.20 \times 10^{-3}$ & 23288.43 & $4.63\times 10^{9}$ & 1000 W & Bacterial Presence\\
\hline
\multicolumn{2}{|c|}{Exposure Time 5 minutes} & \multicolumn{5}{|c|}{}\\
\hline
1 & \emph{B. Subtilis} & $1.0 \times 10^{-3}$ & 35174.24 & $4.67\times 10^{9}$ & 1000 W & Bacterial Presence\\
\hline
2 & \emph{B. Subtilis} & $1.0 \times 10^{-3}$ & 35174.24 & $4.67\times 10^{9}$ & 1000 W & Bacterial Presence\\
\hline
3 & \emph{B. Subtilis} & $1.0 \times 10^{-3}$ & 35174.24 & $4.67\times 10^{9}$ & 1000 W & Bacterial Presence\\
\hline
4 & \emph{B. Subtilis} & $1.0 \times 10^{-3}$ & 35174.24 & $4.67\times 10^{9}$ & 1000 W & Bacterial Presence\\
\hline
\multicolumn{2}{|c|}{Exposure Time 30 minutes} & \multicolumn{5}{|c|}{}\\
\hline
1 & \emph{B. Subtilis} & $2.75 \times 10^{-3}$ & 29624.01 & $1.71\times 10^{9}$ & 1000 W & Bacterial Presence\\
\hline
2 & \emph{B. Subtilis} & $2.75 \times 10^{-3}$ & 29624.01 & $1.71\times 10^{9}$ & 1000 W & Bacterial Presence\\
\hline
3 & \emph{B. Subtilis} & $2.75 \times 10^{-3}$ & 29624.01 & $1.71\times 10^{9}$ & 1000 W & Bacterial Presence\\
\hline
4 & \emph{B. Subtilis} & $2.75 \times 10^{-3}$ & 29624.01 & $1.71\times 10^{9}$ & 1000 W & Bacterial Presence\\
\hline
5 & S. Thermo & $2.75 \times 10^{-3}$ & 28682.98 & $1.81\times 10^{9}$ & 1000 W & Bacterial Presence\\
\hline
6 & S. Thermo & $2.75 \times 10^{-3}$ & 28682.98 & $1.81\times 10^{9}$ & 1000 W & Bacterial Presence\\
\hline
7 & S. Thermo & $2.75 \times 10^{-3}$ & 28682.98 & $1.81\times 10^{9}$ & 1000 W & Bacterial Presence\\
\hline
8 & S. Thermo & $2.75 \times 10^{-3}$ & 28682.98 & $1.81\times 10^{9}$ & 1000 W & Bacterial Presence\\
\hline
\end{tabular}
\end{table*}

\begin{table*}[!ht]
    \centering
    \caption{Petri dish samples results}\label{T2}
    \renewcommand{\arraystretch}{1.5}
     \begin{tabular}{|c|c|c|c|c|c|c|c|}
\hline
Sample & Exposure Time (mins) & Bacteria & Pressure & Temperature (K) & Density (cm$^{-3}$) & Microwave Power & Survivability of spores\\
\hline
1 & 10 & S. Thermo & $2.31 \times 10^{-3}$ & 29526.84 & $5.23\times 10^9$ & 1000 W  & No Bacterial Presence\\
\hline
2 & 5 & S. Thermo & $2.62 \times 10^{-3}$ & 32705.72 & $5.78\times 10^9$ & 1000 W & No Bacterial Presence\\
\hline
3 & 1 & S. Thermo & $2.60 \times 10^{-3}$ & 29662.9 & $6.96\times 10^9$ & 1000 W & Bacterial Presence\\
\hline
4 & 10 & S. Thermo & $3.71 \times 10^{-2}$ & 28817.58 & $3.00\times 10^9$ & 500 W  & Bacterial Presence\\
\hline
5 & 5 & S. Thermo & $2.40 \times 10^{-3}$ & 30798.12 & $3.08\times 10^9$ & 500 W  & Bacterial Presence\\
\hline
6 & 1 & S. Thermo & $1.0 \times 10^{-3}$ & 25976.41 & $3.73\times 10^9$ & 500 W  & Bacterial Presence\\
\hline
7 & 10 & \emph{B. Subtilis} & $2.31 \times 10^{-3}$ & 32561.21 & $6.40\times 10^9$ & 1000 W & No Bacterial Presence\\
\hline
8 & 5 & \emph{B. Subtilis} & $4.31 \times 10^{-2}$ & 31304.23 & $6.44\times 10^9$ & 1000 W & No Bacterial Presence\\
\hline
9 & 1 & \emph{B. Subtilis} & $1.96 \times 10^{-2}$ & 30478.68 & $6.07\times 10^9$ & 1000 W & No Bacterial Presence\\
\hline
10 & 10 & \emph{B. Subtilis} & $3.40 \times 10^{-2}$ & 26205.2 & $5.20\times 10^8$ & 500 W  & Bacterial Presence\\
\hline
11 & 5 & \emph{B. Subtilis} & $2.30 \times 10^{-2}$ & 37471.12 & $9.05\times 10^8$ & 500 W  & Bacterial Presence\\
\hline
12 & 1 & \emph{B. Subtilis} & $4.0 \times 10^{-2}$ & 144097.3 & $6.42\times 10^8$ & 500 W  & Bacterial Presence\\
\hline
\end{tabular}

\end{table*}

\section{Discussion and Results}

\begin{figure}
  \centering
  \includegraphics[width=.8\columnwidth]{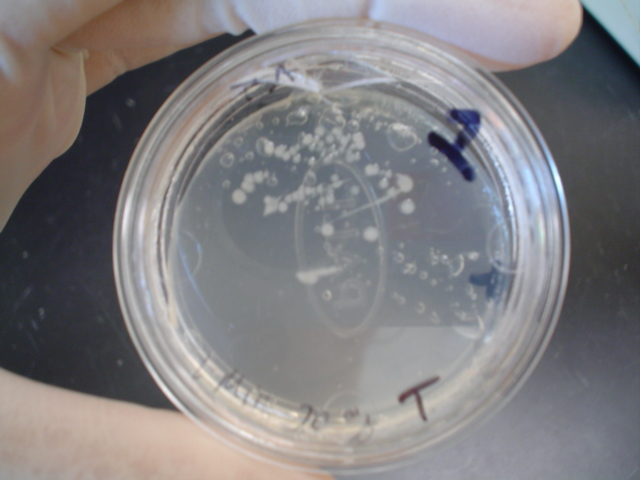}
  \caption{Positive result, \emph{B. Stearothermophilus} 1000 W MW, 1 minute.}\label{F4}
\end{figure}

\begin{figure}
  \centering
  \includegraphics[width=.8\columnwidth]{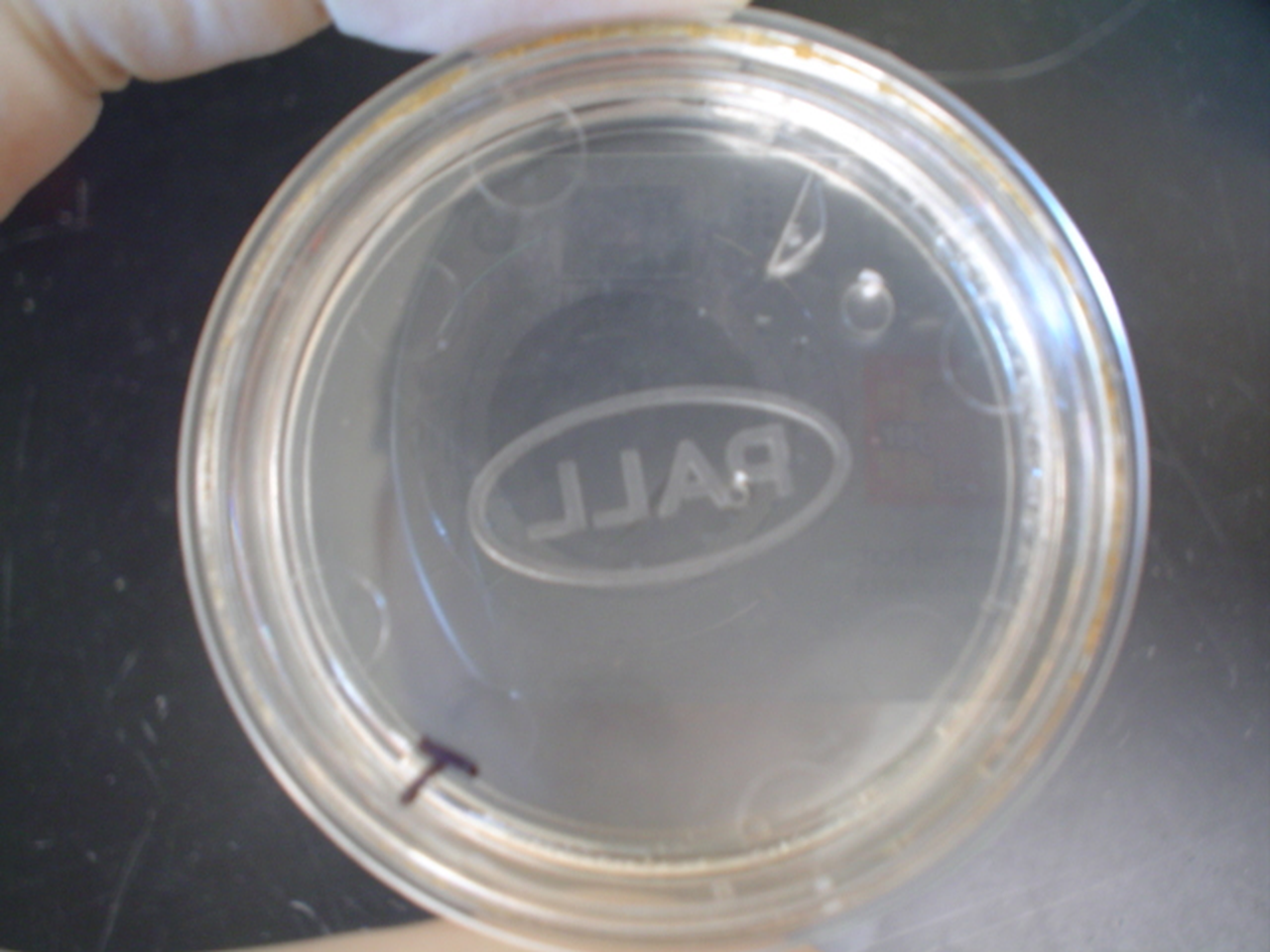}
  \caption{Negative result, \emph{B. Stearothermophilus} 1000 W MW, 5 minutes.}\label{F5}
\end{figure}

Plasma inactivation is characterized by the existence of two or three distinct phases in the survival curves of a given type of spores \cite{moisan2002plasma}. These phases correspond to changes in the dominating kinetics of spore inactivation as a function of exposure time. In the case of medium and low-pressure discharges, the basic process is the DNA destruction of the spore by UV irradiation, preceded in some cases by erosion of the microorganism through intrinsic photodesorption and etching (eventually enhanced by UV radiation). These elementary mechanisms clearly set plasma sterilization apart from all other sterilization methods. In that respect, the inactivation of pathogenic prions, which have no genetic material, could certainly be achieved through full erosion of these proteins; however, it could be that intense UV radiation, acting synergistically with oxygen atoms, causes substantial damage to these proteins, thereby avoiding the need for full erosion. More recently, we have shown that a better understanding of the inactivation process is attained by using the concept of fluence (number of photons actually interacting with the genetic material over a given period of time) that depends on the absorption coefficient and penetration depth of photons in the spore material.
Fluence is actually more appropriate than the incident flux of UV photons since DNA inactivation results from a given (statistical) number (or dose) of damaging hits on its strands. Efficient plasma sterilization at reduced pressure in O\textsubscript{2} containing mixtures under flowing afterglow conditions depends essentially on maximization of the UV emission intensity. Similar results are expected when the microorganisms are directly exposed to the corresponding discharge. Etching in presence of oxygen atoms provides a faster erosion of the protecting layers of microorganisms than simply photodesorption, only the latter being possible with Hg lamps. Furthermore, sterilization using UV lamps and lasers suffers from shadowing effects (which includes the difficulty to sterilize in crevices), in contrast to gas plasma sterilization where the UV photon is brought to the appropriate site by its emitting atom or molecule. An important shortcoming of plasma sterilization is its dependence on the actual “thickness” of the microorganisms to be inactivated since the UV photons need to reach the DNA. Any material covering the microorganisms, including packaging, will slow down the process.
For the Low temperature, atmospheric pressure plasmas have been shown to possess very effective germicidal characteristics. Their relatively simple and inexpensive designs, as well as their nontoxic nature, give them the potential to replace conventional sterilization methods in the near future. This is a most welcome technology in the healthcare arena where reusable; heat sensitive medical tools are becoming more and more prevalent.
For the oxygen plasma \cite{moreira2004sterilization}; were employed to test their sterilization potential on cellulose strips which contained \emph{Bacillus Stearothermophilus}. When the samples were inserted into a test tube and in this way not directly exposed to the plasma, a partial inactivation occurred. However, when the samples were directly exposed to the oxygen plasma, a complete inactivation took place after only 7min of exposure. SEM analysis showed how the spores were deformed and destroyed. Following the atomic oxygen line with an emission spectrograph, it was possible to detect the end point of the sterilization process, what enables the real time determination of the optimal plasma processing time.
For the survivability of spore there are two important findings in various experiments. First, it was demonstrated that a spore’s degree of survivability in the harsh environment of air plasma is directly dependent on the integrity of its protective coat. Because of their low water contents, spores can resist elevated temperatures effectively. However, in air plasma, the oxygenbased radicals offer an aggressively oxidative environment where the proteins of the spore coat can be denatured. As the coat loses its integrity, the core of the spore becomes vulnerable to attack by the plasma generated radicals. This ultimately can lead to the inactivation of the spores. Second, it was demonstrated that cells that survive exposure to air plasma emerge with measurable metabolic changes. These heterotrophic modifications are indications that enzyme activities may have been affected. It is, however, still not clear if these changes (or for that matter possible changes in other biochemical pathways) are temporary or permanent or if they ultimately lead to death of the cells \cite{laroussi2006spores}.
For the sterilization using low pressure direct current discharge with hydrogen peroxide plasma the accomplished experiments show that: the main role in sterilization of open surfaces by plasma of the discharge in hydrogen peroxide solution vapor is performed by active particles formed in the plasma, not by UV radiation of the plasma \cite{soloshenko2002features}. Sterilization by hydrogen peroxide plasma is more efficient than sterilization by plasma of the discharge in oxygen, especially for packed articles and articles with complex shapes. 
As it can be observed in Table 1, there was bacterial presence in all the samples even setting the microwave power to 1000 W (30 volts or 990 watts) and an exposure time of 30 minutes. 
Now, for the results tabulated in Table 2, it can be seen that there was no bacterial presence for the samples that were exposed to oxygen plasma with a microwave power of 1000 W except for the \emph{B. Stearothermophilus} that was exposed for 1 minute (Figure \ref{F4}). For the \emph{B. Subtilis}, all of the samples exposed to oxygen plasma with microwave power of 1000 W showed no bacterial presence at all. But for all the other samples that were exposed to plasma with 500 W  of microwave power a bacterial presence was observed. When looking at the results, it can be seen that with the decrease in microwave power a decrease in plasma density and electron temperature was observed. These decreases may be the source of the positive results in bacterial growth.
With these observations it can be said that all of the objectives were achieved. First, the experimental method was designed or else no experiment could be done; second, the parameters determined to be the most effective were: for the \emph{B. Stearothermophilus}, oxygen plasma with microwave power of 1000 W and a time of exposure of at least 5 minutes (Figure \ref{F5}); and for the \emph{B. Subtilis}, oxygen plasma with microwave power of 1000 W and an exposure time of at least 1 minute. Third, the most effective method was to put the samples in petri dishes rather than in tubes; it can be said that the difference in results was possibly because of the distance of the bacteria to the plasma. In the tube samples the bacteria was not directly in contact with the plasma, but in the Petri dish samples they were exposed in a more direct manner.

\section{Conclusion}

From these experiments it can be concluded that all the objectives were achieved; two experimental method designs were compared in order to determine which was the most effective, which resulted in the exposure of the samples in petri dishes for it was concluded that the bacteria were more directly exposed to the oxygen plasma rather than in the tube samples. Also with the results obtained certain parameters of effectiveness were determined; for the \emph{B. Stearothermophilus}, it was observed that the sterilization parameters had to be set to a microwave power of 1000 W and an exposure time of 5 minutes, at least. For the \emph{B. Subtilis}, the sterilization parameters determined to be the most effective were a microwave power of 1000 W and an exposure time of at least 1 minute. After exposing the bacteria to the oxygen plasma discharges, the procedure of growing them was executed, when the time for growth of both bacteria ended it was determined that putting the samples in petri dishes was more effective, because at the parameters set to be the more convenient no presence of bacteria was shown.


\begin{thebibliography}{10}
\providecommand{\url}[1]{#1}
\csname url@samestyle\endcsname
\providecommand{\newblock}{\relax}
\providecommand{\bibinfo}[2]{#2}
\providecommand{\BIBentrySTDinterwordspacing}{\spaceskip=0pt\relax}
\providecommand{\BIBentryALTinterwordstretchfactor}{4}
\providecommand{\BIBentryALTinterwordspacing}{\spaceskip=\fontdimen2\font plus
\BIBentryALTinterwordstretchfactor\fontdimen3\font minus
  \fontdimen4\font\relax}
\providecommand{\BIBforeignlanguage}[2]{{%
\expandafter\ifx\csname l@#1\endcsname\relax
\typeout{** WARNING: IEEEtranS.bst: No hyphenation pattern has been}%
\typeout{** loaded for the language `#1'. Using the pattern for}%
\typeout{** the default language instead.}%
\else
\language=\csname l@#1\endcsname
\fi
#2}}
\providecommand{\BIBdecl}{\relax}
\BIBdecl

\bibitem{chen2012introduction}
F.~F. Chen, \emph{Introduction to plasma physics}.\hskip 1em plus 0.5em minus
  0.4em\relax Springer Science \& Business Media, 2012.

\bibitem{goldston2020introduction}
R.~J. Goldston, \emph{Introduction to plasma physics}.\hskip 1em plus 0.5em
  minus 0.4em\relax CRC Press, 2020.

\bibitem{gonzalez2024water}
A.~Gonzalez-Lizardo, G.~Morales, J.~Rondon, J.~Santiago, A.~Landron, and
  X.~Pena, ``Water purification via plasma,'' \emph{arXiv preprint
  arXiv:2405.09524}, 2024.

\bibitem{kumar2021basic}
R.~Kumar and A.~Rana, ``Basic concepts of sterilization techniques,''
  \emph{Research Journal of Pharmacology and Pharmacodynamics}, vol.~13, no.~4,
  pp. 155--161, 2021.

\bibitem{laroussi2006spores}
M.~Laroussi, O.~Minayeva, F.~Dobbs, and J.~Woods, ``Spores survivability after
  exposure to low-temperature plasmas,'' \emph{IEEE transactions on Plasma
  Science}, vol.~34, no.~4, pp. 1253--1256, 2006.

\bibitem{moisan2002plasma}
M.~Moisan, J.~Barbeau, M.-C. Crevier, J.~Pelletier, N.~Philip, and B.~Saoudi,
  ``Plasma sterilization. methods and mechanisms,'' \emph{Pure and applied
  chemistry}, vol.~74, no.~3, pp. 349--358, 2002.

\bibitem{moreira2004sterilization}
A.~J. Moreira, R.~D. Mansano, T.~d. J.~A. Pinto, R.~Ruas, L.~da~Silva~Zambon,
  M.~V. da~Silva, and P.~B. Verdonck, ``Sterilization by oxygen plasma,''
  \emph{Applied surface science}, vol. 235, no. 1-2, pp. 151--155, 2004.

\bibitem{nishikawa2000plasma}
K.~Nishikawa and M.~Wakatani, \emph{Plasma physics}.\hskip 1em plus 0.5em minus
  0.4em\relax Springer Science \& Business Media, 2000, vol.~8.

\bibitem{rutala2019guideline}
W.~A. Rutala and D.~J. Weber, ``Guideline for disinfection and sterilization in
  healthcare facilities, 2008. update: May 2019,'' 2019.

\bibitem{soloshenko2002features}
I.~Soloshenko, V.~Tsiolko, V.~Khomich, V.~Y. Bazhenov, A.~Ryabtsev,
  A.~Schedrin, and I.~Mikhno, ``Features of sterilization using low-pressure
  dc-discharge hydrogen-peroxide plasma,'' \emph{IEEE transactions on plasma
  science}, vol.~30, no.~4, pp. 1440--1444, 2002.

\end{thebibliography}
\end{document}